%% file: eprint.tex
\documentclass[10pt]{article}
\usepackage{graphicx}
\usepackage{lineno}

\def\Title#1{\begin{center} {\Large #1 } \end{center}}
\def\Author#1{\begin{center}{ \sc #1} \end{center}}
\def\Address#1{\begin{center}{ \it #1} \end{center}}

\newcommand\pubblock{\rightline{\begin{tabular}{l} Proceedings of the Fifth Annual LHCP\\ \pubnumber\\
         \pubdate  \end{tabular}}}

\newenvironment{Abstract}{\begin{quotation} \begin{center} 
             \large ABSTRACT \end{center}\bigskip 
      \begin{center}\begin{large}}{\end{large}\end{center} \end{quotation}}

\newenvironment{Presented}{\begin{quotation} \begin{center} 
             PRESENTED AT\end{center}\bigskip 
      \begin{center}\begin{large}}{\end{large}\end{center} \end{quotation}}

\input econfmacros.tex

\usepackage{color}

\newcommand\pubnumber{ ATL-PHYS-PROC-2017-213 }

\newcommand\pubdate{\today}

\def\affiliation{
On behalf of the ATLAS Experiment, \\
Department of Physics \\
Stockholm University, 106 91 Stockholm, Sweden}

\begin{document}


\large
\begin{titlepage}
\pubblock

\vfill
\Title{Top pair production in association with a vector gauge boson in ATLAS}
\vfill

\Author{ J\"orgen Sj\"olin  }
\Address{\affiliation}
\vfill
\begin{Abstract}

An overview of the latest results for top pair production in association with a vector gauge boson in the ATLAS detector at LHC is presented. The results
involving $Z$ and $W$ bosons are recorded at $\sqrt{13}$ TeV collision energy,
while the results involving photons are recorded at $\sqrt{7}$ TeV and
$\sqrt{8}$ TeV collision energy.

\end{Abstract}
\vfill

\begin{Presented}
The Fifth Annual Conference\\
 on Large Hadron Collider Physics \\
Shanghai Jiao Tong University, Shanghai, China\\ 
May 15-20, 2017
\end{Presented}
\vfill
\end{titlepage}
\def\thefootnote{\fnsymbol{footnote}}
\setcounter{footnote}{0}
%

\normalsize 


\section{Introduction}

Interactions in the top quark sector are of central interest at the LHC since many new physics scenarios involve top quarks. One example of an important class of scenarios are those that address the Higgs hierarchy problem. For models residing at high energy scales this induces new electroweak top interactions that can be accurately parametrized using effective field theory. Top quark production in association with vector bosons is a strong experimental handle on the leading effective operators that preserve charge-parity and flavour in neutral currents. The experimental input for setting effective theory limits are precision measurements of inclusive and differential cross sections. The latest results from the ATLAS detector~\cite{Aad:2008zzm} in this class of measurements are presented in the following sections.

\section{Top pairs in association with a $Z$ or $W$ boson}
\begin{figure}[htb]
\centering
\includegraphics[width=10cm]{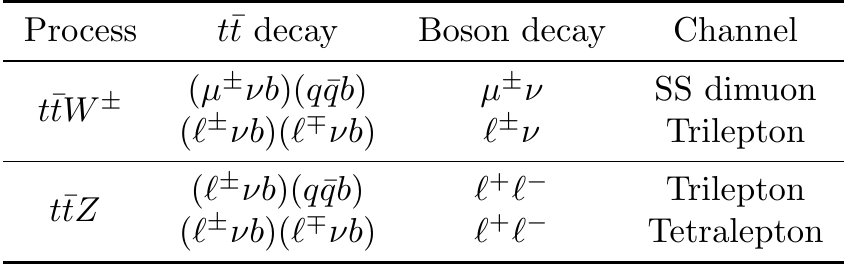}
\caption{List of leading significant final states included in the presented
$t\bar{t}Z$ and $t\bar{t}W$ analysis.}
\label{fig:table1}
\end{figure}

\begin{figure}[htb]
\centering
\includegraphics[width=13cm]{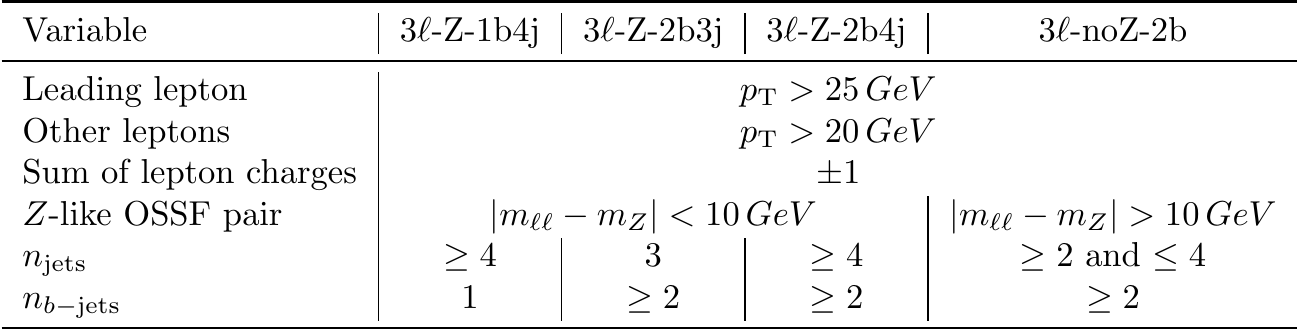}
\caption{Summary of event selections in the trilepton signal regions.}
\label{fig:table2}
\end{figure}

\begin{figure}[htb]
\centering
\includegraphics[width=13cm]{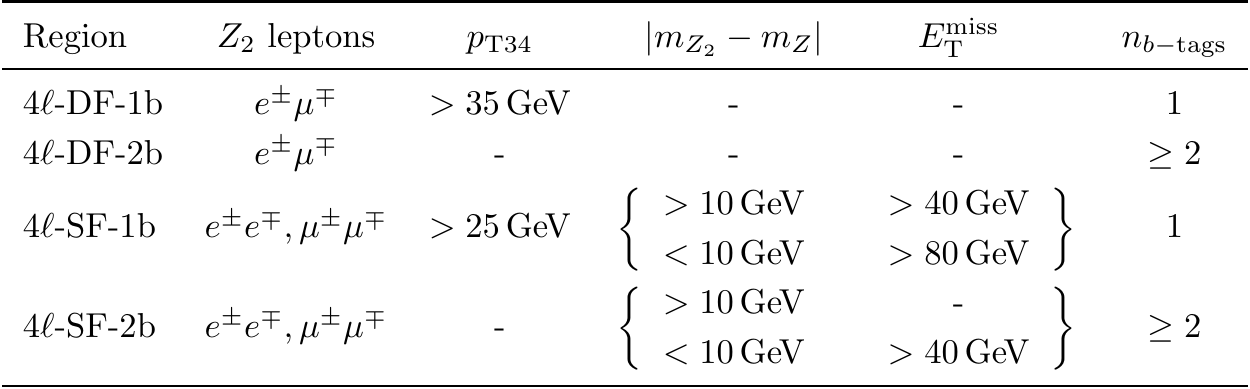}
\caption{Summary of event selections in the tetralepton signal regions.}
\label{fig:table3}
\end{figure}

Top quark pairs in associations with heavy vector bosons produce a significant
signal in several final states involving leptons, see Figure \ref{fig:table1}
for the leading significant examples involving $Z$ and $W$ bosons.
To enhance the sensitivity, each final state is optimized using cuts and then
combined to extract the common cross sections using a profile likelihood fit.
The resulting cuts from the optimization are shown in Figure \ref{fig:table2}
for trileptons and in Figure \ref{fig:table3} for tetraleptons.
The same-sign dimuon events are selected by requiring the muon transverse
momementum $p_T>25$ GeV, missing transverse momentum $E^{miss}_T>40$ GeV, the
scalar sum of $p_T$ of leptons and jets ($H_T$) larger than 240 GeV, and at
least two b-tagged jets.

\begin{figure}[htb]
\centering
\includegraphics[width=4cm]{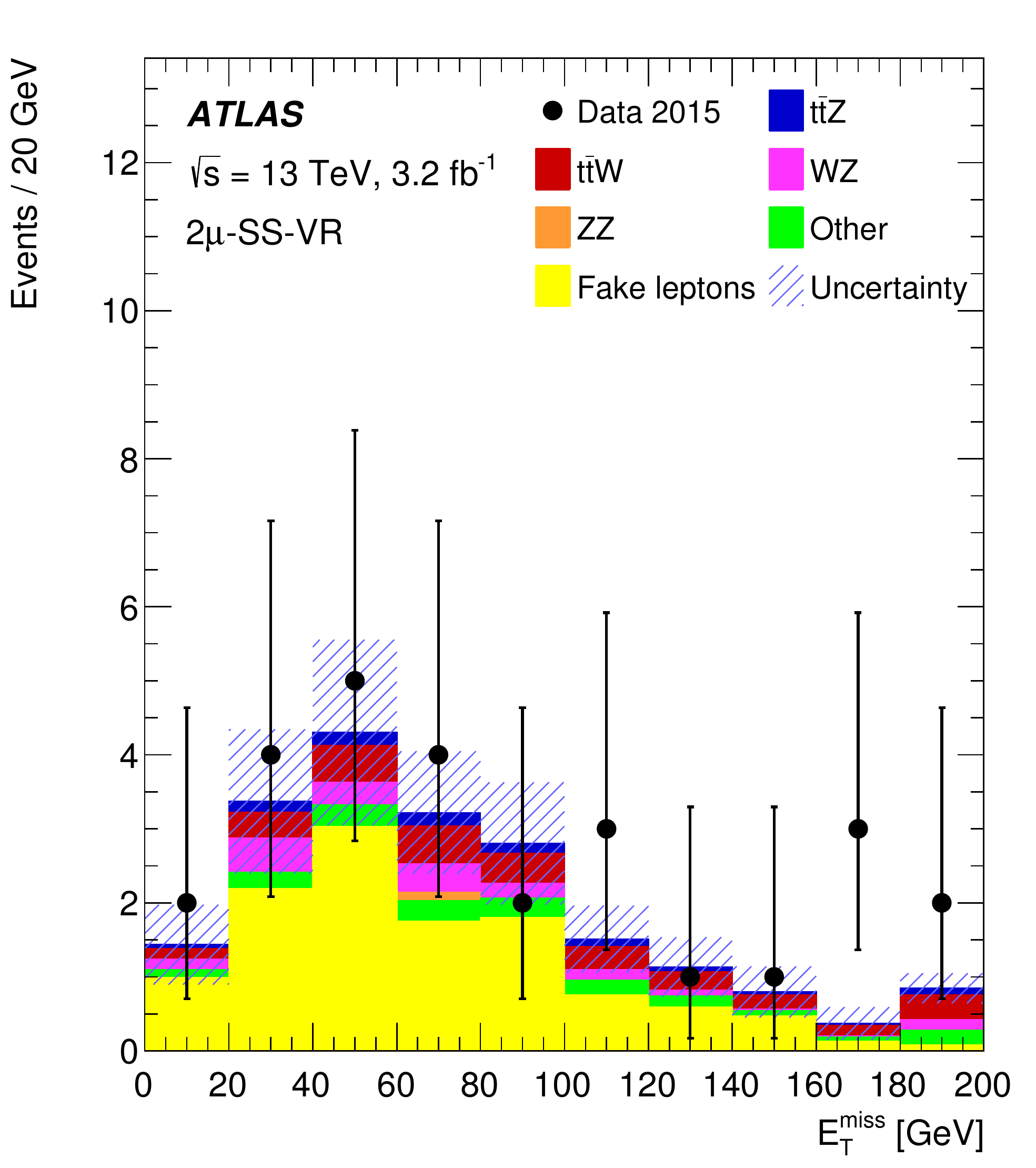}
\includegraphics[width=4cm]{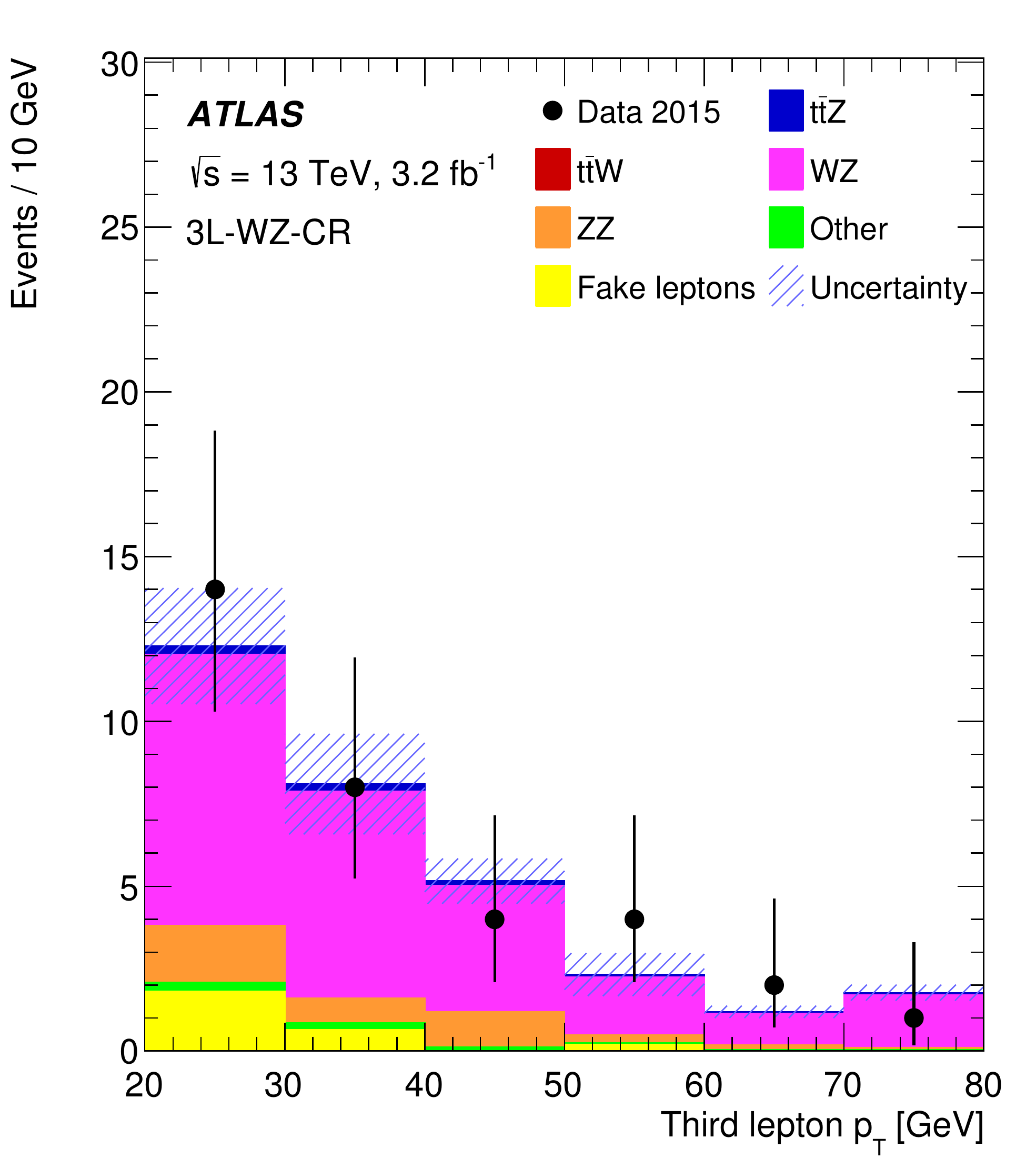}
\includegraphics[width=4cm]{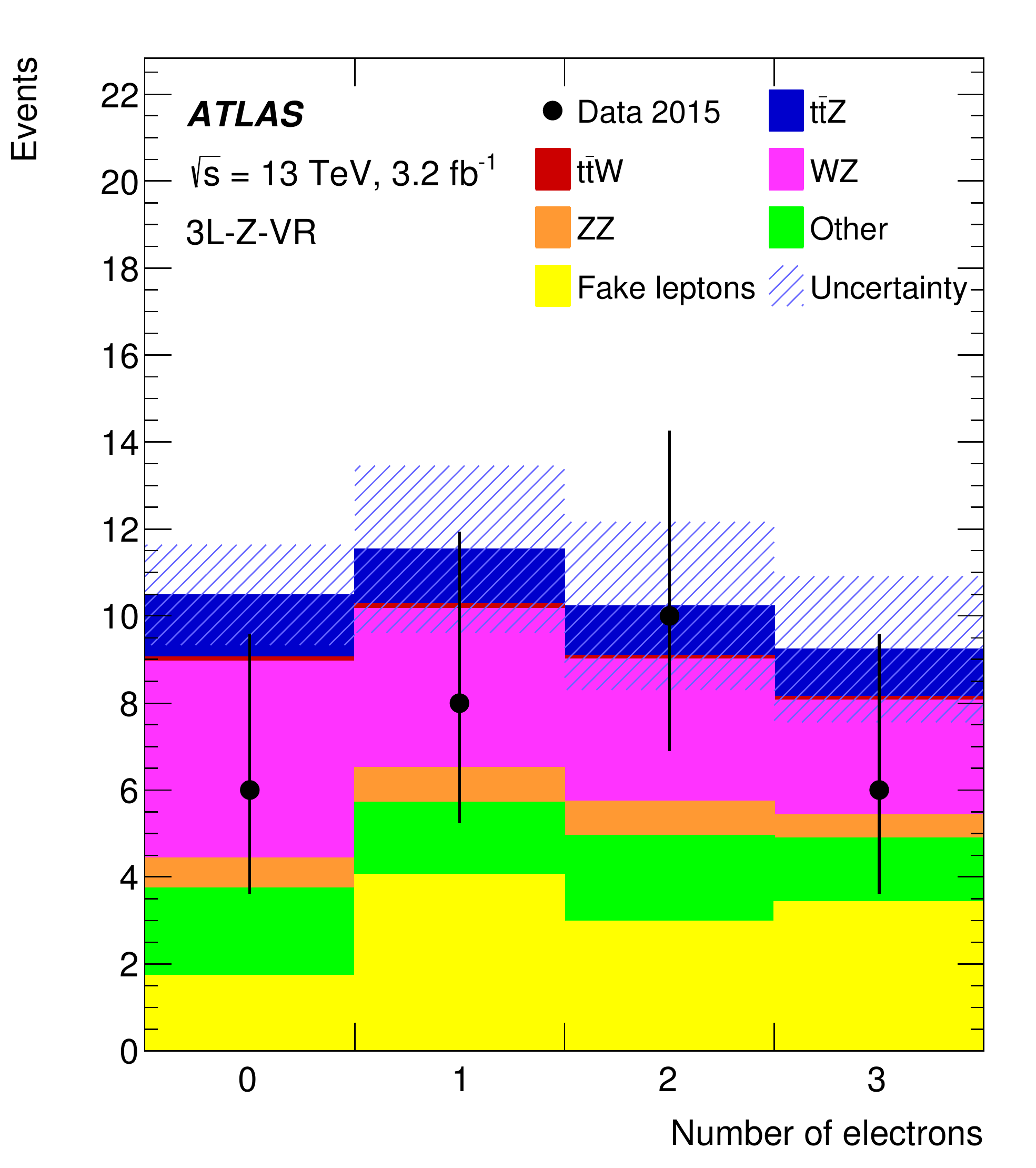}
\includegraphics[width=4cm]{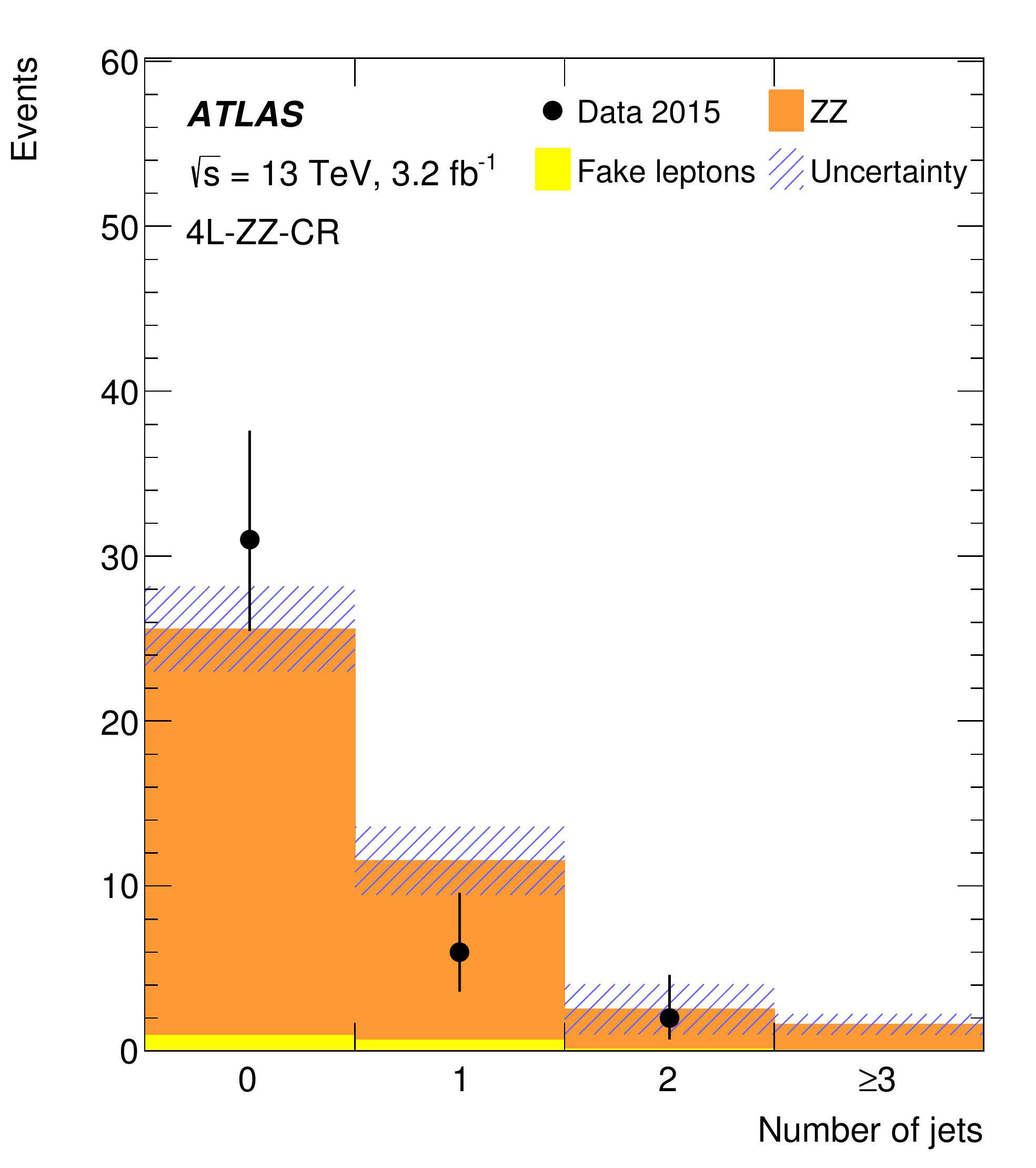}
\caption{Examples of background validation in the $t\bar{t}(W/Z)$
measurement~\cite{Aaboud:2016xve}.
From the left: fakes and non-prompt enhanced event distribution of $E^{miss}_T$
in same-sign muons, third leading lepton $p_T$ in a $WZ$ enhanced trilepton
region, lepton flavor distribution in a trilepton region close to the signal
region, and jet multiplicity in a $ZZ$ enhanced tetralepton region.}
\label{fig:vr}
\end{figure}

The dominating backgrounds vary between the final states. For same-sign
leptons non-prompt and fake leptons dominate, in trileptons the $WZ$
background is large and has significant systematic uncertainties due to
extrapolations
into the high jet multiplicity signal regions, while in tetraleptons the
$tWZ$ and $ZZ$ dominate. The diboson backgrounds are estimated by extrapolating
the yields from lower jet multiplicity using Monte Carlo, while non-prompt and
fake leptons are estimated with the Matrix Method. The Matrix Method parameters
are fitted using a Matrix Method likelihood in data control regions, taking
into account lepton $p_T$ and b-tagging multiplicity. Examples of background validation is shown
in Figure \ref{fig:vr}.

\begin{figure}[htb]
\centering
\includegraphics[width=7cm]{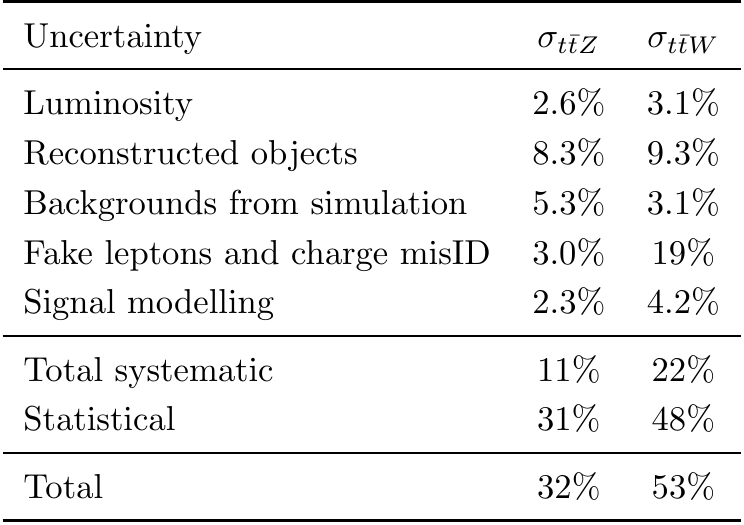}
\caption{Overview of the systematic uncertainties affecting the
$t\bar{t}(W/Z)$ measurement~\cite{Aaboud:2016xve}.}
\label{fig:sys}
\end{figure}

\begin{figure}[htb]
\centering
\includegraphics[width=8cm]{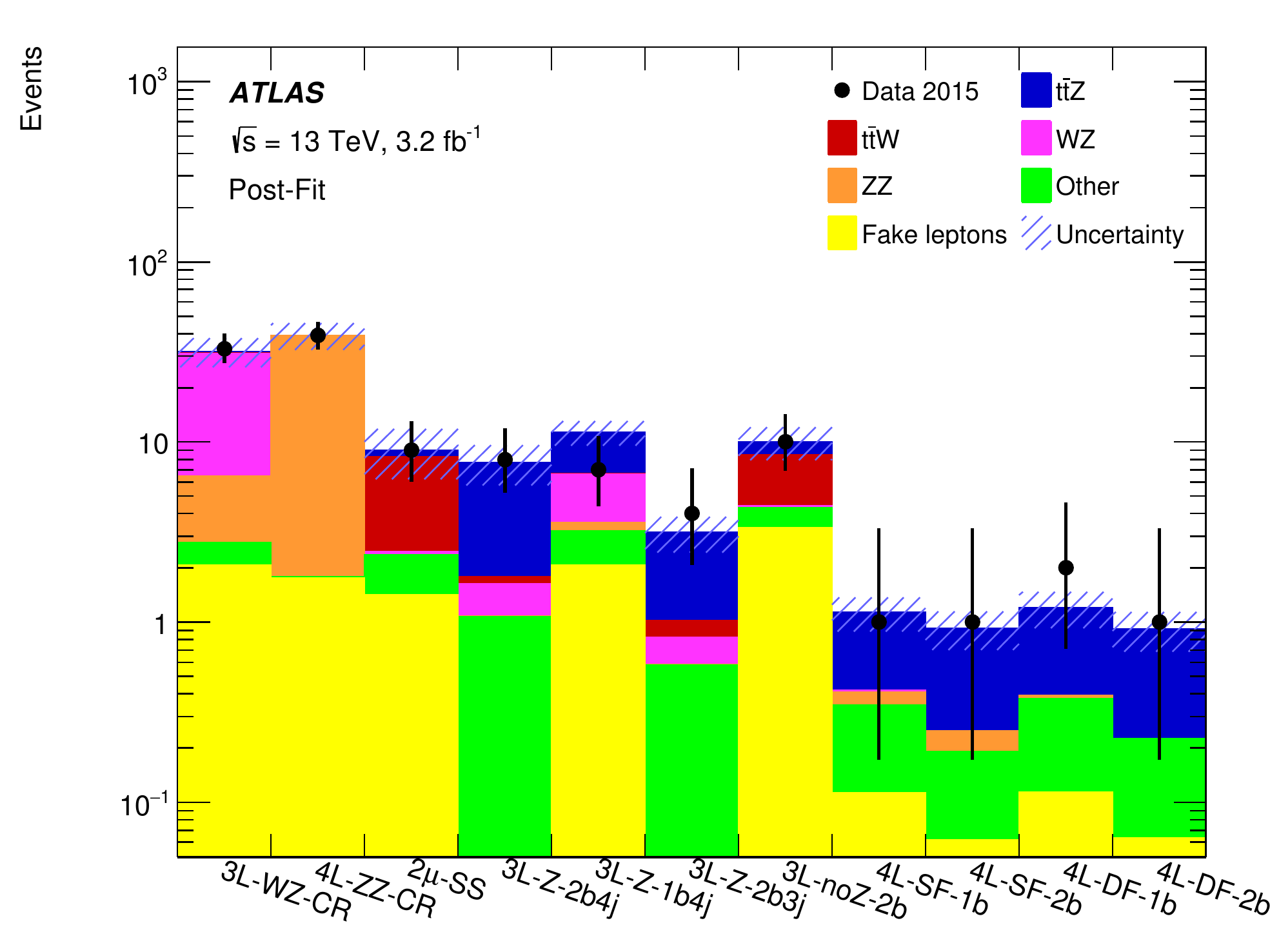}
\includegraphics[width=7cm]{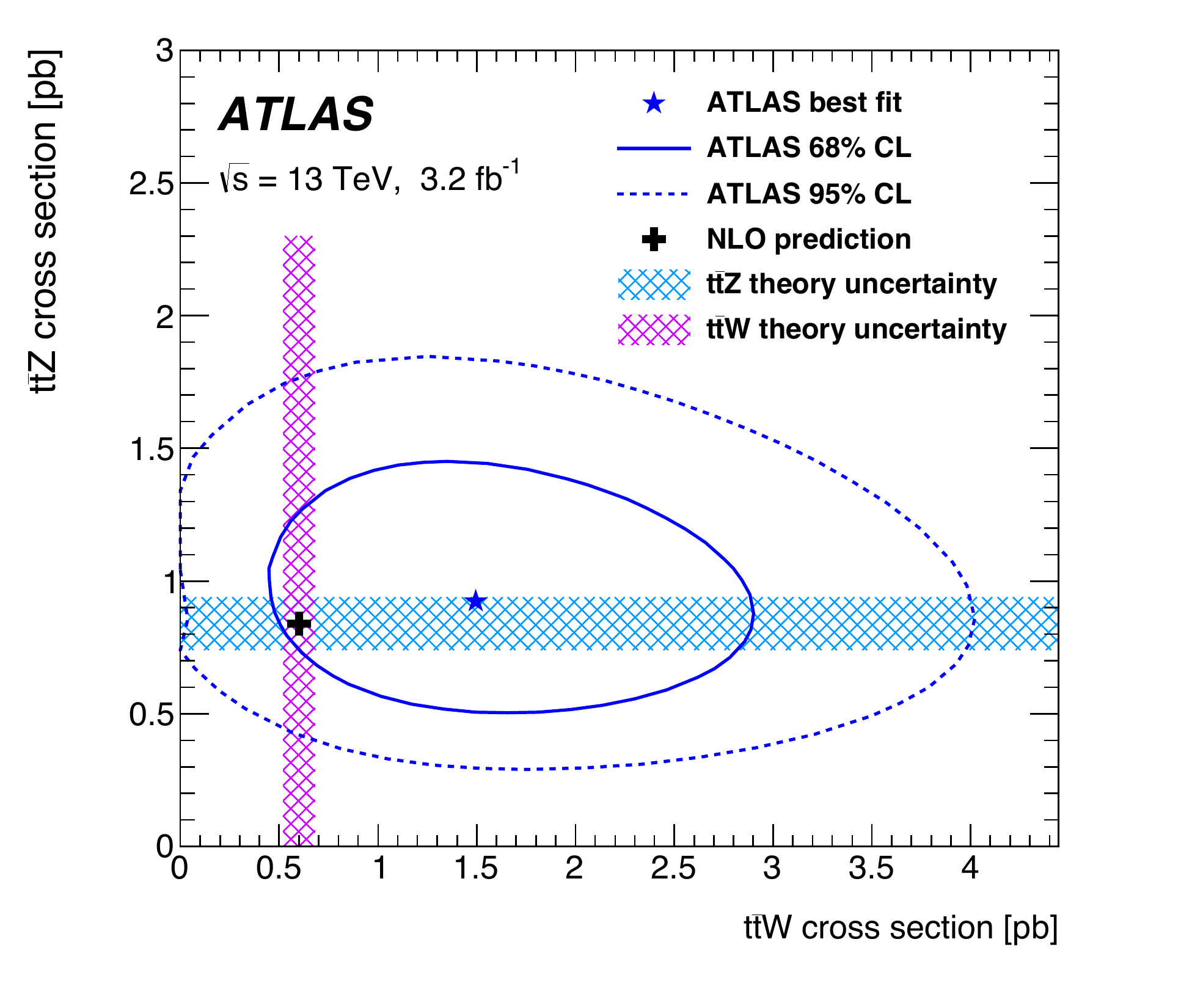}
\caption{The left plots shows a comparison of the yields in data to expectation in the different regions used in the $t\bar{t}(W/Z)$ measurement after the fit.
On the right a comparison is made between the Standard Model cross section
and the simultaneously measured cross sections of $t\bar{t}W$ and $t\bar{t}Z$.
The plots are taken from Ref.~\cite{Aaboud:2016xve}.}
\label{fig:fit}
\end{figure}

An overview of uncertainties affecting the cross section measurement is listed in Table \ref{fig:sys},
and the measured yields and estimated cross sections are shown in Figure \ref{fig:fit}.

\section{Top pairs in association with a photon}
ATLAS has performed measurements of top pairs in association with a photon using data collected at $\sqrt{s}=7$ TeV, see Ref. \cite{Aad:2015uwa}.
Preliminary updated and improved results are also available for
data collected at $\sqrt{s}=8$ TeV, see Ref. \cite{Aaboud:2017era}.
The measurements are performed in a fiducial volume of the detector with
photon $p_T>20$ GeV (7 TeV data) and photon $p_T>$ 15 GeV (8 TeV data).

The main backgrounds originate
from both prompt and non-prompt contributions, and the fractions are
determined using a template fit to the photon track isolation within a
$\Delta R \le 0.2$ cone around the photon. Depending
on the process both data and simulations are used to estimate the
isolation template shapes. 

The prompt photon isolation distribution of the signal is
estimated using electrons from $Z$ boson decays, corrected both for the
difference between electrons versus photons and the topology difference between
$Z$ decays and $t\bar{t}\gamma$ using simulations, in
bins of $p_T$ and pseudo-rapidity ($\eta$). 

\begin{figure}[htb]
\centering
\includegraphics[width=8cm]{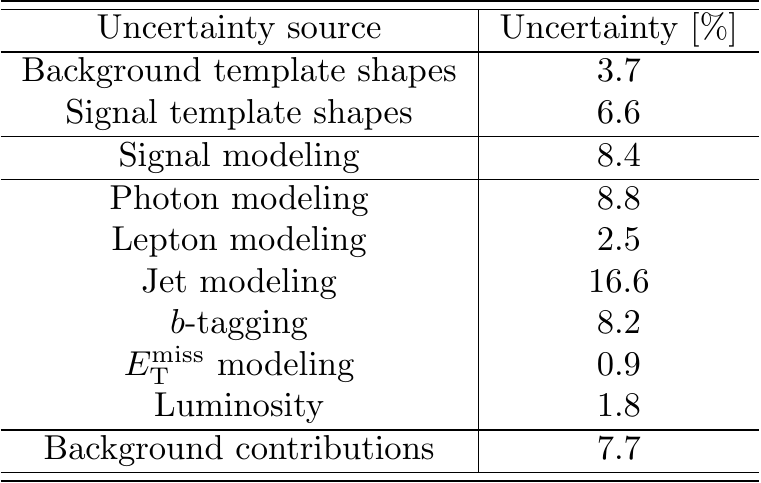}
\caption{Overview of the systematic uncertainties affecting the
7 TeV data $t\bar{t}\gamma$ measurement~\cite{Aad:2015uwa}.}
\label{fig:ttg_sys}
\end{figure}

\begin{figure}[htb]
\centering
\includegraphics[width=7cm]{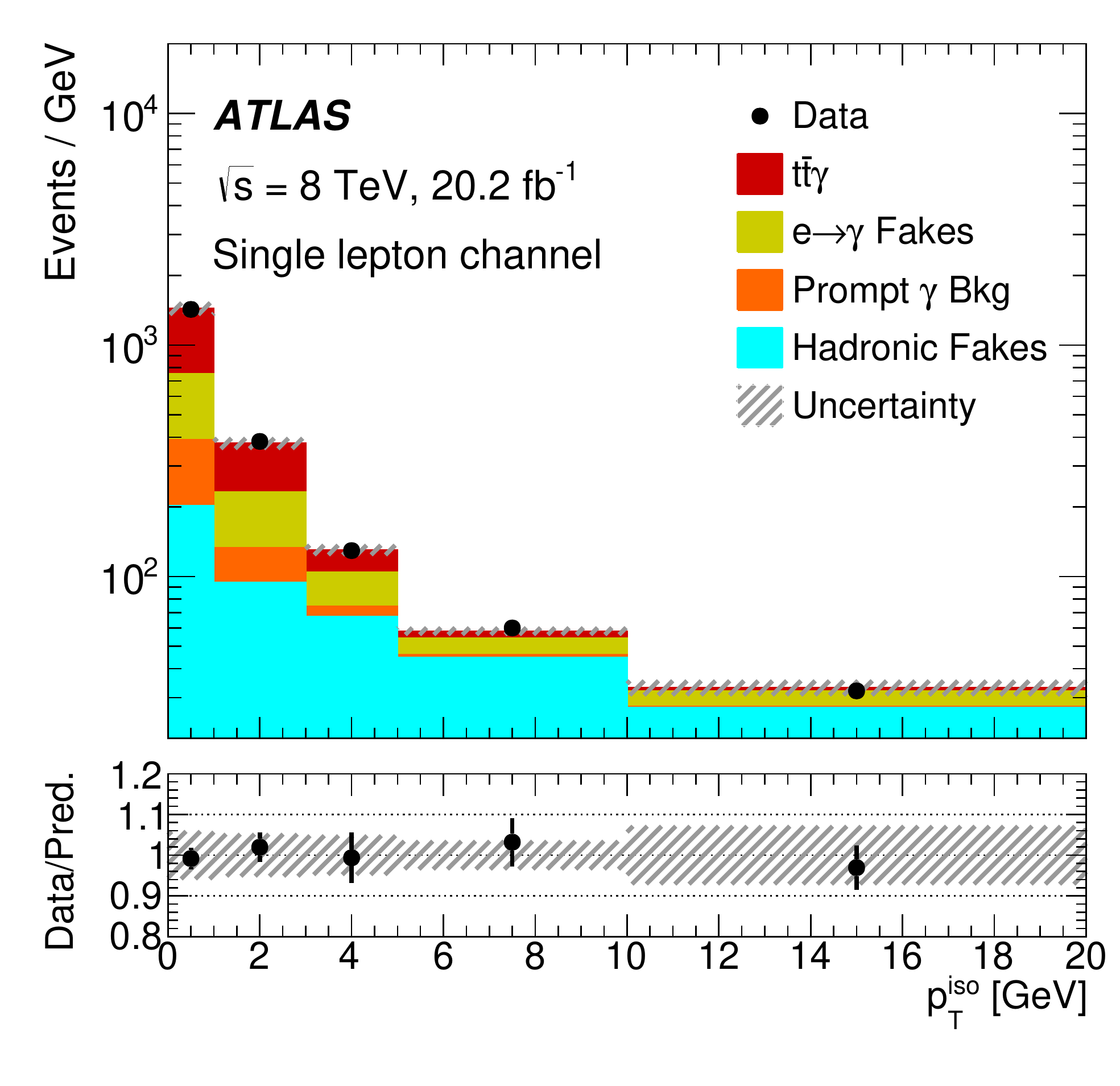}
\includegraphics[width=7cm]{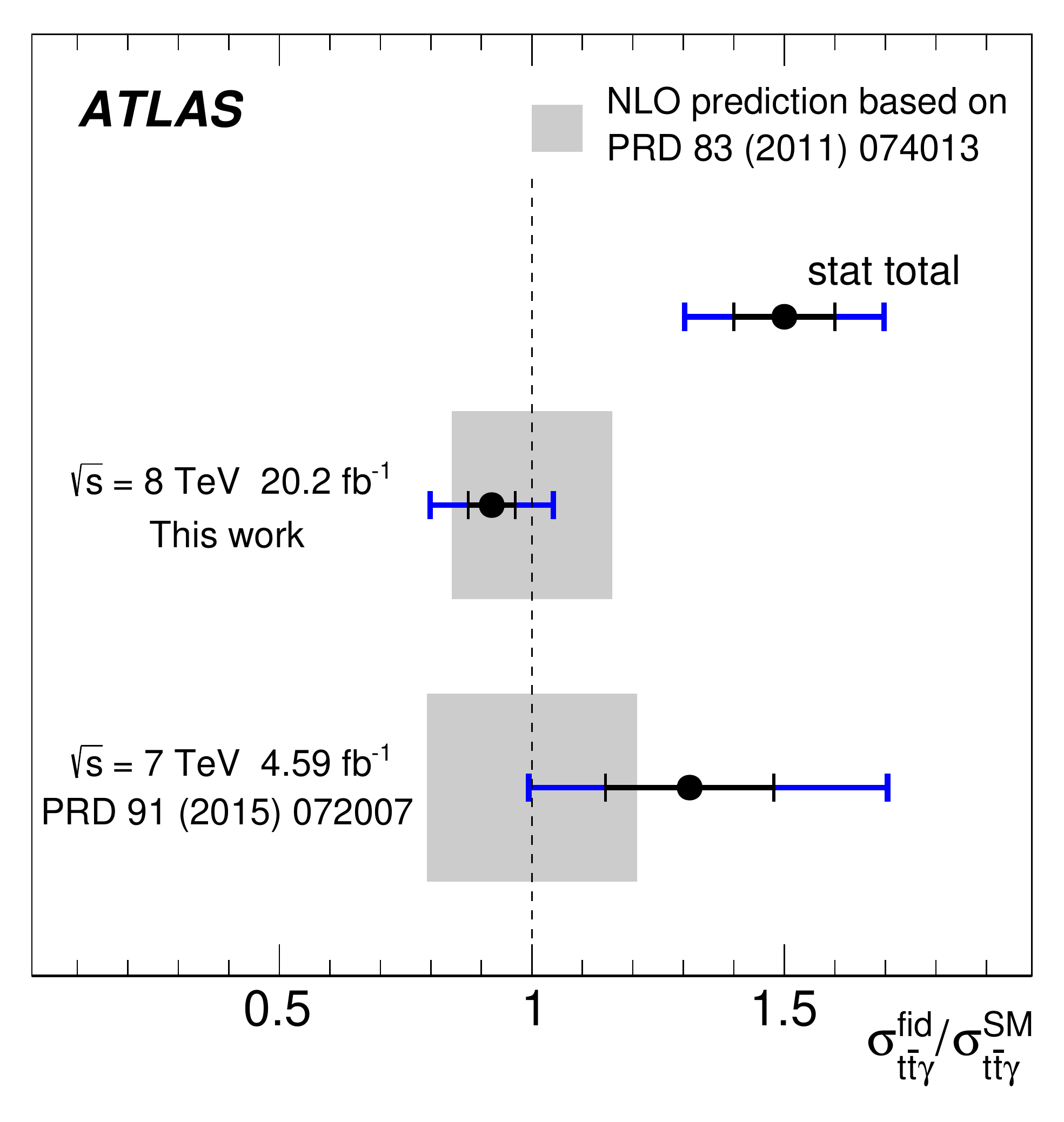}
\caption{The left plot shows the estimated signal and background fractions after
fitting the photon track isolation to data. On the right a comparison is made
between the Standard Model prediction of the $t\bar{t}\gamma$ inclusive
cross section and the data measurements, both for the 7 TeV
data measurement~\cite{Aad:2015uwa} and the updated 8 TeV data
measurement~\cite{Aaboud:2017era}.}
\label{fig:ttg_inc}
\end{figure}

\begin{figure}[htb]
\centering
\includegraphics[width=7cm]{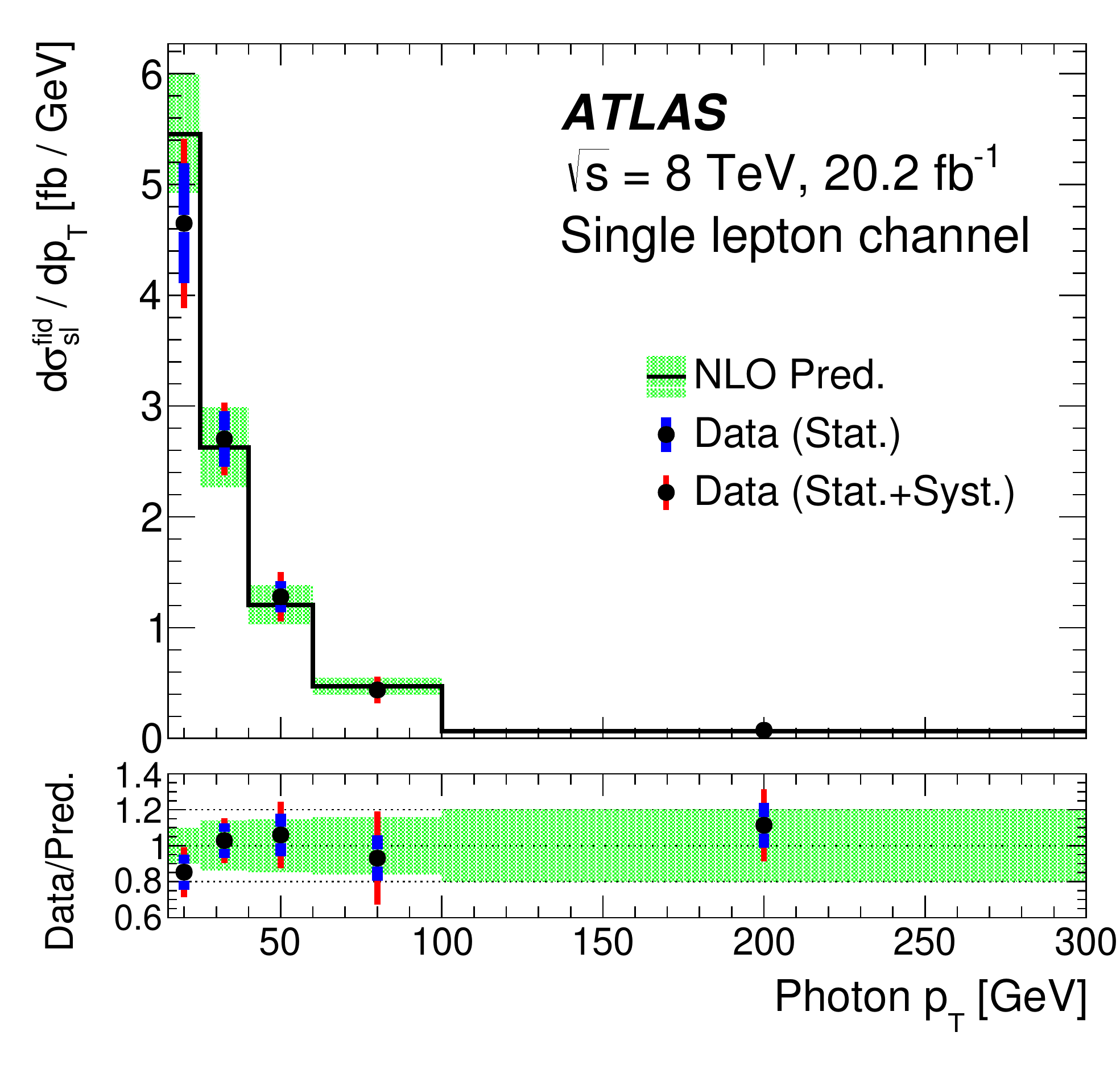}
\includegraphics[width=7cm]{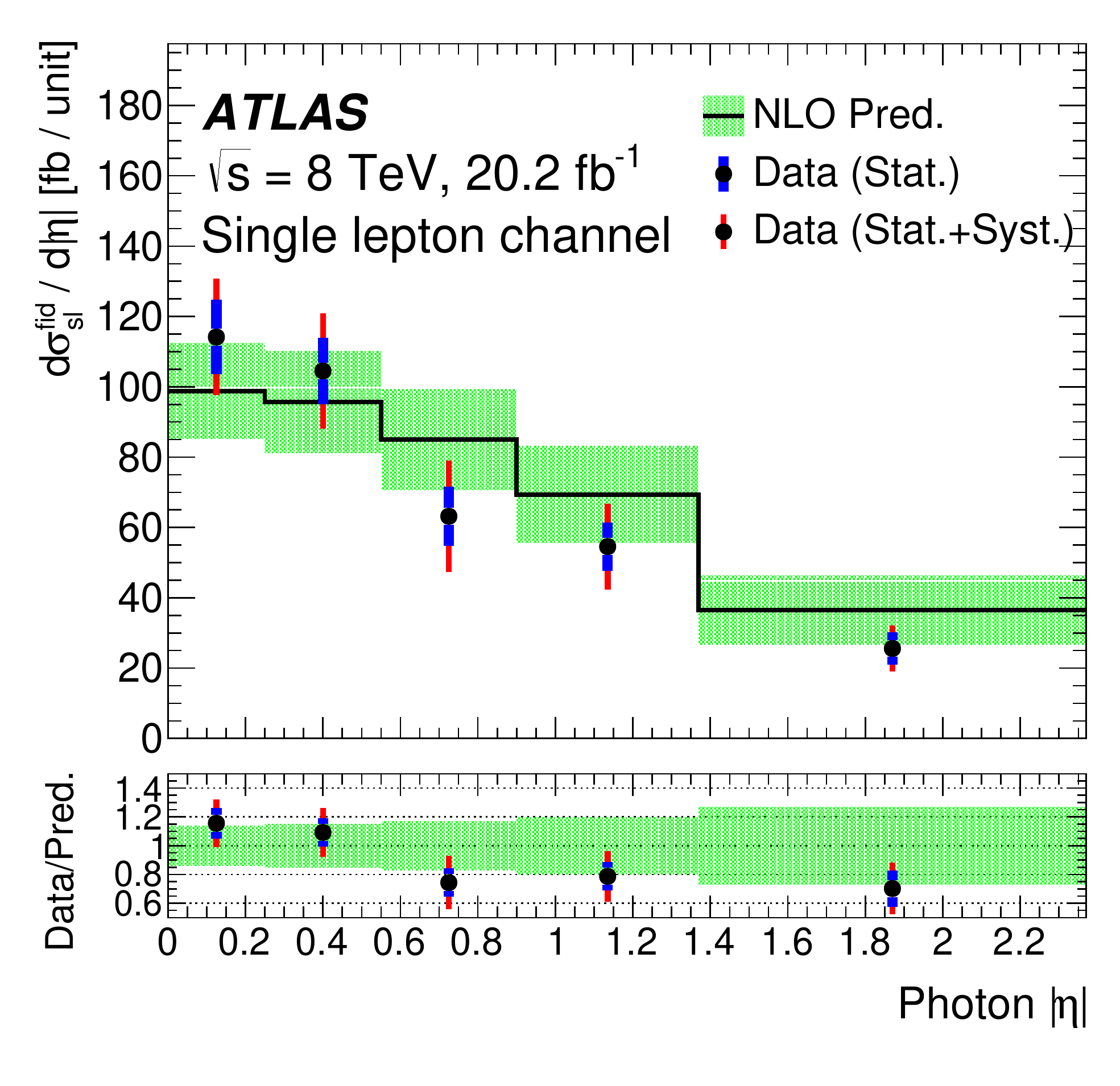}
\caption{The plots show the detector unfolded differential distributions
of the photon $p_T$ (left) and photon $\eta$ (right) estimated in the
8 TeV data $t\bar{t}\gamma$ measurement~\cite{Aaboud:2017era}.}
\label{fig:diff}
\end{figure}

Contrary to the $t\bar{t}W$ and $t\bar{t}Z$ measurements, the $t\bar{t}\gamma$
measurements are limited by systematic uncertainties, see Table \ref{fig:ttg_sys} for an
overview of the different sources. The measured
fractions from signal and different backgrounds after the fit to data
are shown in Figure \ref{fig:ttg_inc}.
In the updated 8 TeV data measurement, a measurement is also made of the
photon differential distributions in $p_T$ and $\eta$, unfolded for detector
effects. The differential distributions are shown in Figure \ref{fig:diff}.

\section{Conclusions}

The production of top pairs with associated bosons is a very active research
field at the LHC. The strong interest is partly prompted by the strong
ability to constrain EFT operators in the top sector.

Published ATLAS results have been shown for associated $W$, $Z$ and $\gamma$
production as well
as some new preliminary results of the associated $\gamma$ production.
The focus in the updated measurements are towards making the measurements
less model dependent (using well defined fiducial volume) and provide more
information beyond just inclusive cross sections, i.e.
differential cross sections.

Currently all measured cross sections agree well with the Standard Model
predictions. However, the performance of the results are limited by
precision, and consequently this
kind of measurements will become more and more important in the future as
higher statistics LHC data samples become available.


\end{document}

%% file: econfmacros.tex
\def\beq{\begin{equation}}
\def\eeq#1{\label{#1}\end{equation}}
\def\eeqn{\end{equation}}

\def\beqa{\begin{eqnarray}}
\def\eeqa#1{\label{#1}\end{eqnarray}}
\def\eeqan{\end{eqnarray}}

\let\bar=\overbar

\def\Dslash{\not{\hbox{\kern-4pt $D$}}}
\def\dslash{\not{\hbox{\kern-2pt $\del$}}}

\def\msb{{\bar{\ssstyle M \kern -1pt S}}}

